*New Update, March, 2015:*

# Fakes as Likes and Dislikes as Likes:
## A Comprehensive Study of the Flaws of the Facebook Like System

Cyber Physical System Lab
School of Computer Science
McGill University

The initial work of this project was conducted from September 2012 to December 2012 in the Cyber Physical System Laboratory in the School of Computer Science at McGill University. By December 2012, we had identified a series of flaws associated with the design and implementation of the Facebook Like system. We reported these flaws to Facebook in February 2013, and expressed our intention to collaborate on helping fix them. The Site Integrity Team of Facebook replied in March 2013 acknowledging the inherently insecure design, but they need to spend more engineering time than research collaboration. After that, we submitted some of the research findings to one of the most famous convention – DEFCON, under the title of "Are You Really Liking It When You Use the Facebook Likes" in May 2013, and the paper got accepted in late June 2013. Later, because we could not go to present at the conference, we had to retract the paper from the conference.

Over the past two years, Facebook has made continuous progress fighting with fake likes. Facebook released patches and improvements to the Like button API and fixed some of the major flaws. In the mean time, the Like button has become the de facto standard for the users to show their fondness of a particular online content, and been used as a widely-accepted metric to measure the popularity of a webpage. Even more interesting, it has been associated with the economic benefits and interests to the providers of the contents / underlying business. There are reports stating that a single Facebook like is worth as much as $174 to the business [1, 2].

However, we recently found that several of the flaws we discovered are still out in the wild. We recorded 3 demo videos to illustrate these flaws and potential threats. For example, these flaws could be used by online spammers to generate massive amount of fake Facebook Likes for profits, impeding the common interest of both social network users and legitimate advertisers. This may endanger the ecosystems on the Internet which leverages on the Facebook API. Also, we show these fake "likes" can be easily generated in an automatic fashion at very low cost. As another example, we show some of the flaws also lead to the increase of Facebook Likes even when legitimate users are making negative comments and expressing dislikes (or even disgusts) of the associated online content. These findings surprise most people.

We discover that a large number of the online websites, including famous ones like Yahoo, abcNews, HuffingtonPost, FoxNews, ESPN, BillBoard, etc., are affected by these flaws. Given the fact that Facebook has become an integral part of our everyday digital lives, and is the dominant online forum for social networking, these flaws may have potentially large negative impacts and consequences to our online community. We all love Facebook. We hope by making public our research findings together with several video demos can help raise the awareness of these flaws and potential consequences. We hope these efforts can contribute to the research and solutions of building a truthful and healthy online social ecosystem.

These demos, which were recorded in March, 2015, are available at:
Demo 1: https://youtu.be/lB9KKDRQ52c
Demo 2: https://youtu.be/LffXRnEzdV0
Demo 3: https://youtu.be/lxcW75FKCoM

[1] David Cohen, Syncapse: Each Facebook Like Is Worth $174 To Brands, retrieved on March, 4[th], 2015 from
http://www.adweek.com/socialtimes/syncapse-like-174/418690
[2]. Courtney Kettmann, Is a Facebook "Like" Worth $174? Probably Not, retrieved on March, 4[th], 2015 from
http://www.wired.com/2013/07/is-a-facebook-like-worth-174-probably-not/

# Does "Like" Really Mean Like?

## A Study of the Facebook Fake Like phenomenon and an efficient countermeasure

Xinye Lin,    Mingyuan Xia,    Xue Liu

{xinye.lin, mingyuan.xia}@mail.mcgill.ca
xueliu@cs.mcgill.ca
School of Computer Science
McGill University
Montreal, Canada

## ABSTRACT

Social networks help to bond people who share similar interests all over the world. As a complement, the Facebook "Like" button is an efficient tool that bonds people with the online information. People click on the "Like" button to express their fondness of a particular piece of information and in turn tend to visit webpages with high "Like" count. The important fact of the Like count is that it reflects the number of actual users who "liked" this information. However, according to our study, one can easily exploit the defects of the "Like" button to counterfeit a high "Like" count. We provide a proof-of-concept implementation of these exploits, and manage to generate 100 fake Likes in 5 minutes with a single account. We also reveal existing counterfeiting techniques used by some online sellers to achieve unfair advantage for promoting their products. To address this fake Like problem, we study the varying patterns of Like count and propose an innovative fake Like detection method based on clustering. To evaluate the effectiveness of our algorithm, we collect the Like count history of more than 9,000 websites. Our experiments successfully uncover 16 suspicious fake Like buyers that show abnormal Like count increase patterns.

## 1. INTRODUCTION

The modern society has witnessed the rapid expansion of social networks, which connect people and deliver huge amount of information everyday. As one of the largest social networks, Facebook has over one billion monthly active users [11]. The Facebook "Like" button is the fundamental mechanism to measure the popularity of certain content on the Internet. With the popularity of Facebook, more and more websites are integrating this button into their webpages. According to a statistics [5], over 3 million websites are using this button today.

The Facebook Like button, as shown in Fig. 1, can be embedded into essentially any webpage and appear also extensively in the Facebook website itself (e.g. a Facebook post, a comment). The Like button has a web button and a count number associated. The number conceptually represents the amount of Facebook users that have clicked on the button to express their fondness of the content. Thus, a higher number indicates that the content is more popular. Since this idea is coined, Facebook has collected more than 1 trillion clicks from its users [11], and the Like button becomes an important tool for online advertising [20, 22].

The great success of Facebook Likes lies in the fact that the count number can precisely represent the number of people that like this content, thus other users could easily find out what is popular by checking the Facebook Like numbers. This is also true for online advertiser. The precision of the number enables advertisers to advertise on popular web pages that generate more values. If the number cannot precisely capture the popularity of the content, it will endanger the whole Facebook ecosystem. In this paper, we reveal the flaws in the Facebook like mechanism and show how these deficiencies are exploited by spammers to generate fake "Likes".

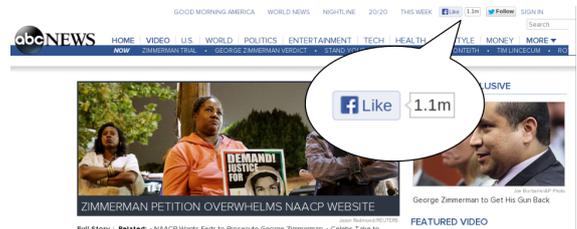

Figure 1: An example of the Facebook Like button

According to our study, online spammers have already been selling huge number of Facebook Likes. These Likes are **intentionally** generated for unfair profits (e.g. for advertisement revenues), which will harm the Facebook community. Fig. 2 presents a search for such services on eBay, which returns more than 3,000 results fake Like sellers. We also find similar sales on fiverr.com and many other freelance websites. There are even companies particularly selling Facebook Likes, such as WeSellLikes [30] and BuyRealLikes [6].

In this paper, we provide a comprehensive study on the Facebook Like mechanism and reveal the flaws that are exploited for counterfeiting likes. We present a proof-of-concept approach, which can generate more than 20 fake Likes per minute with only one spammer Facebook account. To help prevent fake Likes, we propose a clustering-based algorithm to sift out highly suspicious fake Like buyers. We use the algorithm to analyze traces of Like number fluctuations of 9,000 websites. Our results, as validated by manual inspection, effectively uncover the suspicious fake Like buyers.

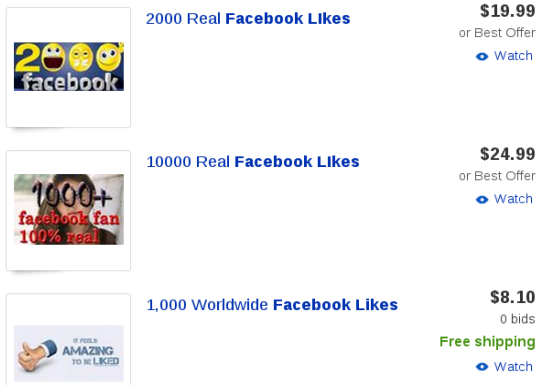

Figure 2: Facebook Like sales on eBay

The contribution of this paper is three-fold:

1. We uncover the fake Like problem and reveal concrete exploits that can generate fake Likes quickly and stealthily (against most known anti-spamming techniques).

2. We propose a unsupervised clustering-based algorithm to sift out highly suspicious fake Like buyers.

3. We present a real world study of 9,000 Alexa top websites using our algorithm, and sucessfully detect 16 suspects of fake Like buyers.

This paper is organized as follows. Section 2 elaborates the Like mechanism, its deficiencies and a proof-of-concept fake Like generator; Section 3 describes our algorithm that identifies suspects and our evaluation results; Section 4 introduces related work; Finally, Section 5 concludes this paper and discusses the future work.

## 2. FAKE LIKES

In this section, we first elaborate the Like mechanism from Facebook. Then, we analyze and reveal potential deficiencies within the current design. Finally, we provide several proof-of-concept methods to generate fake Likes by exploiting these deficiencies. We also compare our methods with existing approaches from massive spamming community.

### 2.1 Design of the Like button

The Facebook Like mechanism is generally used to measure and distribute the popularity of certain webpages on the Internet. This mechanism involves interactions between multiple parties. A *website* embeds the "Like" component provided by *Facebook*. The "Like" component consists of a button and a count number by its side, as shown in Fig. 1. A *user*, upon visiting the webpage, can click on the button to express fondness of the page. When the button is pressed, it reports to Facebook with the URL of the page. Facebook updates the count for the particular URL and next time the page is liked (possibly by other users), the count number will increase by one.

Besides clicking on the button, other activities related to a URL (e.g. sharing the URL as shown in Fig. 3) on the Facebook website will also increase the count for that page. We will explain further in the following section.

An *attacker*, however, disguises as a normal user and abuses the Like button to counterfeit lots of Likes. These so-called fake Likes can fool normal users into believing that the webpage has a great popularity. The attacker achieves this by exploit the flaws in the counting scheme. We will then explain the counting scheme and then dive into the flaws.

### 2.2 Counting Details

#### 2.2.1 Count Classification

As mentioned before, the web component from Facebook contains a button and a number. Before looking into how this number is counted, we define the following terms for better distinguishment:

- The **display count** is the number shown beside the Like button.

- The **like count** is the number of actual clicks on the button. A user can only click once on the button embedded for the same page. The second click will invalidate the first one.

- The **share count** represents the number of times that this page (its URL) is shared on the Facebook website.

- The **comment count** represents the number of comments that appear under the Facebook post that shares the URL.

According to the official Facebook documentation [13], the display count is stored as "total_count" in the database, and it is the sum of the other three above-mentioned counts, i.e.,

$$total\_count = share\_count + like\_count + comment\_count \quad (1)$$

In the rest of this paper, we use `X-count` as a general term to refer to these counts.

#### 2.2.2 Counting Scheme

We perform a series of experiments to reveal the effects of various activities. The activities we looked into comprise common activities on the Facebook website. Tab. 1 presents the summary results. We notice that some activities are repeatable given one user account (test 2, 4, and 6). This provides a potential opportunity for an attacker to generate fake Likes with only one account.

#### 2.2.3 Deficiencies

By examining tests in Tab. 1, we summarize the following design deficiencies which could be exploited.

**Duplicated counts.** As mentioned before, the second click on the button will invalidate the first click. This simply prevents duplicated Likes. However, we found that an attacker can create multiple posts with the same page (test 3) and share these posts for multiple times (test 4) to increase the `share count` of the page, which in turn increases the display count. Similarly, by adding random comments under the post can arbitrarily increase the `comment count` (test 6). These repeatable activities from the same account will result in arbitrary increase in display count.

**Lack of accountability.** Facebook maintains a list of people that like a page. Thus each `like count` is uniquely linked to one user. Furthermore, each `share count` and each `comment count` can link back to a share post or a comment, respectively. This bookkeeping allows Facebook to gain accountability for each count, which gives critical evidence for defending against spamming.

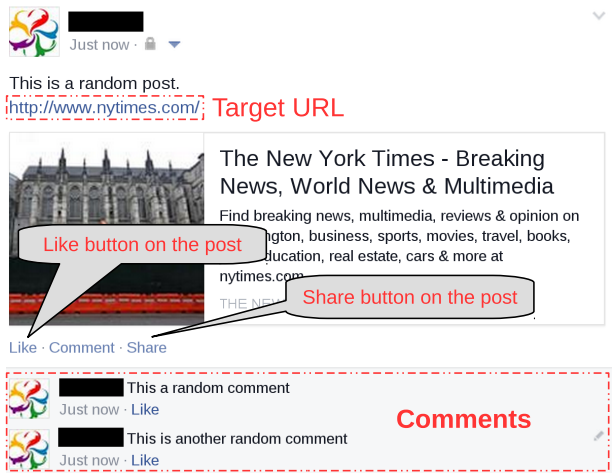

Figure 3: An example of the Facebook post. Test 2 and 4 in Tab.1 correspond to click on the like and share button above the comment area respectively.

However, according to test 5, deleting a post, hence the removal of the comments belonging to it, will not result in any reduction for `share count` and `comment count`. To confirm this, we first create a webpage on our own web server, then share the page on Facebook and make random comments, then immediately delete the post. The display count is then examined everyday for a month. We also carefully clear the web cache and try on several different machines. We conclude that the deletion does not cause any decreasing effect on the display count.

This deficiency essentially cripples the accountability of `share count` and potentially `comment count`, which makes it hard for Facebook to discover spamming posts and accounts.

**Irrelevant counts.** According to the official Facebook documentation [13], "a single click on the Like button will 'like' pieces of content on the web and share them on Facebook". This reflects a common user expectation, that the display count is the number of clicks on the button, which corresponds to `like count`. However, the display count comprise not only `like count` but also `share count` and `comment count`. In test 3, creating random Facebook posts with a given URL will increase `share count` of that URL, even if the posts are expressing dislikes. This is similar for comments. Experiences of other social networks suggest that, a "dislike" button can potentially distinguish these situations and cause less confusion to users. For example, Youtube provides such "dislike" button, Quora and StackExchange give users the "downvote" option, etc.

## 2.3 Generate Fake Likes

We give several proof-of-concept methods to exploit revealed deficiencies to generate fake Likes. By combining and extending these basic generating methods, the attacker can produce massive fake Likes efficiently.

### 2.3.1 Pre-requisite

In the following demonstration, we only register one Facebook account to generate fake likes. We generate a simple webpage on our web server with a Like component from Facebook. The webpage can be replaced by any other webpages on the Internet.

### 2.3.2 Facebook's Anti-spamming

In designing a practical and automated spamming method, we need to understand the anti-spamming mechanisms. We identify the following mechanisms that could affect fake like generation:

**Post spam filter.** The post spam filter will detect posts or comments with identical content and remove them. Furthermore, based on the accountability, an account that generates such duplicated posts or comments will be labeled as spammer and also removed.

**Rate control.** Facebook provides open APIs, which can be used to create, delete and manipulate posts and comments. Any automated fake like generation needs to utilizes these APIs. However, these APIs have rate control, which limits both the frequency and the quantity of API requests. A user that always exceeds the rate control could be labeled as spammer and removed. Yet, an attacker can learn this control policy with several dry runs.

### 2.3.3 Efficient Fake Liking

Based on the deficiencies we revealed, we repeat the below four operations to generate fake likes.

1. Create a Facebook post with the target URL.
2. Share the post just created.
3. Add a comment to the post with arbitrary commenting content.
4. Delete the post.

This procedure can generate three fake likes a time [1]. By repeating it, we manage to generate 20 fake likes per minute (without violating Facebook's rate limit). Also, we delete the post just created to break the accountability, which disables Facebook's spam filter effectively and makes the generation stealthy. We achieve these with only one account. We expect that with more accounts, the generation process could be done much faster. The code is implemented with public Facebook API with 20 lines of python code.

### 2.3.4 Alternative Spamming Methods

We compare our fake like generation process with some well-understood spamming techniques.

**Visual deception.** These approaches [16] simply replace the Facebook component with a fake image or overlay an high count image over the real display count. These approaches cannot cheat users redirected from Facebook (since the display count is kept in the Facebook database). Also, these replaced static image cannot response to the user clicks.

**Spamming accounts.** These methods [4] require a large number of zombie accounts and generate one Like per account, similarly to "bot-net" attacks. These approaches rely on a method to register a large set of zombie accounts, which could be time consuming. Also, these spamming accounts can be easily detected and removed by the spam filter, with a lot of well studied anti-spamming methods [4, 8, 27].

---

[1] As in March, 2015, this procedure now only generates two likes, because Facebook has removed the flaw which makes creating a post with the target URL generate one Like.

Table 1: Tests and corresponding effects on Facebook Likes

| Tests | | Effects | | |
| --- | --- | --- | --- | --- |
| No. | Details | `like count` | `share count` | `comment count` |
| 1 | click the Like button on the test page | +1 | 0 | 0 |
| 2 | click the Like button on a post[1] | +1 | 0 | 0 |
| 3 | create a post with target URL (repeatable[2]) | 0 | +1 [4] | 0 |
| 4 | share the post with target URL (repeatable[2]) | 0 | +1 | 0 |
| 5 | delete a post with target URL[3] | 0 | 0 | 0 |
| 6 | add a random comment to the post (repeatable[2]) | 0 | 0 | +1 |
| 7 | delete a comment on the post | 0 | 0 | -1 |

[1] All the posts referred in this table are Facebook posts sharing the target test page's URL address with random words.

[2] The test is exactly repeatable, thus will lead to duplicate Likes.

[3] To ensure that the all-zero effect of this test is not due to any network cache or delayed update in the Facebook database, the effects have been validated each day for one month after the post being deleted.

[4] As tested in March 2015, this effect has been disabled by Facebook, and now it should be 0.

Compared to existing spamming methods, our method has the following advantages:

- Single account.

  Our approach does not need to maintain a pool of zombie accounts, which saves lots of costs and time. Nevertheless, if the attacker does have multiple zombie accounts, they can be used in parallel to speedup the fake Like generation at ease.

- Easy to implement.

  The generation process needs only 20 lines of code to complete the attack and is fully automated. With only a small number of zombie accounts, our approach can effectively generate hundreds of fake Like per minute.

- Anti-anti-spamming.

  Our approach breaks the accountability of the Facebook services and carefully obeys rate control policy. Preventing spams without accountability is difficult. Thus our approach can operate stealthily without any direct evidences in the Facebook system.

- Actual spamming.

  Unlike visual deception, our approach places actual changes of the display count in Facebook database. Thus, even a normal Facebook user can view a high display count of the target URL.

## 2.4 Threat assessment

Before the submission of this paper, we have reported and confirmed these deficiencies with Facebook. In response, Facebook released a few patches to the Like mechanism to eliminate some of the deficiencies we mentioned. However, our investigation shows that some of these deficiencies still persist and widely impact websites which use the Facebook Like button. As in March 2015, a list by the company `BuiltWith` counts over 3 million websites that use Facebook Like button [5]. We further look into the Alexa top site list and find that potential victims include popular websites such as CNN, ABC News, The Huffington Post, The Economist, ESPN, Billboard, etc. Given the wide coverage and large reader population of these websites, we believe the flaws of the Facebook Like button is potentially a large threat to the health of the Facebook ecosystem.

## 3. SPOT THE FAKE LIKE BUYERS

In this section, we dig out the suspicious fake Like buyers by exploring the abnormality in the historical Like counts of real world websites. The basic idea is: First, record the historical tracks of different counts of selected websites over an observation period; second, extract "meaningful" segments, and cluster them into different patterns; third, employ these patterns as the features to cluster the websites; finally, spot the suspicious websites by analyzing the clusters. Before getting into the details of data processing, we briefly introduce the clustering algorithms being used.

### 3.1 Data set

The historical display counts of 9,000 websites are monitored over 53 days, spanning from Nov. 10, 2012 to Jan. 1, 2013. These websites are picked from Alexa top 1 million sites [1] ranked on Nov. 5, 2012. Among them, 2,000 are picked from the top, 2,000 from the middle, and the rest 5,000 from the tail of the 1 million sites. They are referred to as the top, the middle and the tail sites hereafter. Such a picking scheme ensures that the data set comprises different kinds of websites. Intuitively, the websites closer to the top are larger websites with large traffic and are known by more people, they have weaker incentive to buy fake Likes. In contrast, the tail websites have less traffic and are known by less people, thus they have stronger incentive to buy fake Likes and less concern about the risk of their reputation. That is the reason for picking more (i.e., 5,000) websites from the tail of the list.

All details of the display count, i.e. `share count`, `like count`, `comment count` and `total count`, are crawled from Facebook using the graph API. The crawling period is set to 12 hours, i.e. two samples are crawled each day, at noon and midnight respectively. In the end, each website has four *count tracks*, with 106 entries per track.

### 3.2 Data preprocessing

### 3.2.1 Remove unchanged tracks

If a track is never changed, it can neither convey any information about normal website activities, nor represent any potential Like buying behaviors. Thus we remove them from the data set at the first step.

### 3.2.2 Convert to the ratio track (normalization)

The crawled websites have a large variance in the base number of `X-count`. Thus it is meaningless to directly compare their counts with each other. For example, for a website with 1 million `like count`, an increment of 1,000 is probably a normal case. But for a website with only 100 `like count`, such an increment is definitely suspicious. To exclude the influence of different base number and make the counts across websites comparable, the absolute numbers in the *count track* are normalized as follows.

For a certain kind of count, the number at the beginning of the observation is denoted as $a$, and that in the end as $b$. Then the absolute change during the observation period is $|b-a|$. Furthermore, if we denote the $i$th record in the track as $n_i$, the variance ratio upon $n_i$ being crawled is defined as:

$$v_i = \begin{cases} 0 & i=1; \\ \dfrac{n_i - n_{i-1}}{|b-a|} & i=2,3,\ldots,len \end{cases}$$

and *len* is the length of the track. $v_i$ reflects how much a certain sample period has contributed to the increment through the whole observation. Notice that the crawled counts are not monotonically increasing over time, thus it is possible that $v_i$ is negative or has an absolute value larger than 100%. Fig. 4a illustrates the variance ratio track of `share count` for `facebook.com`.

After this step, each website has four *ratio tracks*, which store the variance ratio of `share count`, `like count`, `comment count` and `total count` respectively over the observation period.

## 3.3 Detect suspicious websites

After preprocessing the crawled data, we can start the process of clustering the websites. In this subsection, we first give the basic assumptions, and then elaborate upon how the websites are clustered.

### 3.3.1 Assumptions

According to the observations of the data set and the real world experience, we make the following assumptions before clustering the websites.

**Assumption** 1. *Within each ratio track, at most $L$ consecutive points are related.*

The physical interpretation of the data point in the *ratio track* is how much the `X-count` has increased over the last time it being crawled. In real life, the `X-count` changes in a bubbling fashion: Sometimes the website is promoted by special events, for example, advertised by a celebrity or famous blog, then it gets high increasing ratios for a period, which causes a "bubble". For other time, the variance ratio is stable with minor fluctuations. The time span of the largest bubble defines $L$. In this paper, we choose $L$ to be 5, based on the observation that most Like sales on eBay promise to take effect within 48 hours, given that our crawling period is 12 hours.

**Assumption** 2. *For any two different websites, their ratio tracks are independent from each other.*

There may exist a few cases against this assumption. For example, A and B are websites holding similar content and B has embedded a promotion link of A on its homepage. It is possible that when B gets a significant increase of Likes, the same group of users also visit A and give A Likes. In this case, display count of A and B are not strictly independent. Considering this case, we analyze the source code of the monitored websites for promotion links of each other, and find it to be extremely rare. We conclude that this assumption holds in general and are reasonable for the work in this paper.

**Assumption** 3. *Most of the websites never buy fake Likes.*

This assumption is easy to understand according to the basic rule called "presumption of innocence" in Social Science.

**Assumption** 4. *Top sites are less likely to buy fake Likes.*

We mentioned before that the top sites have larger traffic and are well advertised over the tail sites. They have more concern about the reputation and smaller needs for fake Likes.

### 3.3.2 Extract the website features

According to our observation, the Like count changes of the websites are event-driven. Various real world events, such as festivals, promotions, media reports, etc., lead to the big rises and falls in the recorded ratio tracks. Each event corresponds to a segment of the track comprising a few consecutive ratio points. Segments are independent from each other. Between these segments, the variance ratio are relatively stable with small variances. These segments give hint about the changing patterns of Like counts, thus can be used as the features of websites.

*Abnormal segments.*

To this end, our first goal is to split the *ratio track* into independent segments. According to Assumption 1, the maximal length of the segment is $L$. To find out the delimiting positions between different segments in the *ratio track*, We introduce the *abnormality filter band*. We take the *ratio track* of `facebook.com` to illustrate how it works (Fig. 4).

Denote the *ratio track* of `facebook.com` as $T_r$. The center line of the filter band is the average of $T_r$ (denoted as $\mu(T_r)$, the red dot line in the middle of Fig. 4a); the width of the filter band is twice the standard deviation of $T_r$ (denoted as $\sigma(T_r)$); the upper border and lower border of the filter band are $\mu(T_r)+\sigma(T_r)$ and $\mu(T_r)-\sigma(T_r)$, and correspond to the upper and lower red dot line in Fig. 4a respectively. The filter band confines the "normal zone" of a website. Changing ratios beyond this zone, either higher or lower, are likely to represent real world events.

Then we impose this filter band upon the *ratio track*. For points fall inside this band (eg. $A_1$) or right on the border (eg. $A_2$), we replace them with 0; for points fall outside the band (eg. $B_1$ and $B_2$), we keep them unchanged. After this step, a few consecutive non-zero sequences separated by zero are produced (Fig. 4b). Because they are comprised of points out of the "normal zone", we denote them as the *abnormal segments*.

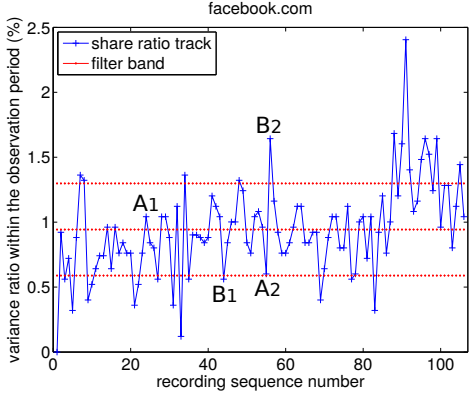 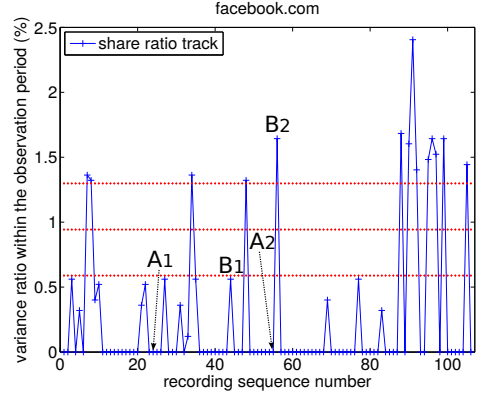

(a) Before      (b) After

Figure 4: An example of extracting *abnormal segments* by applying the filter band.

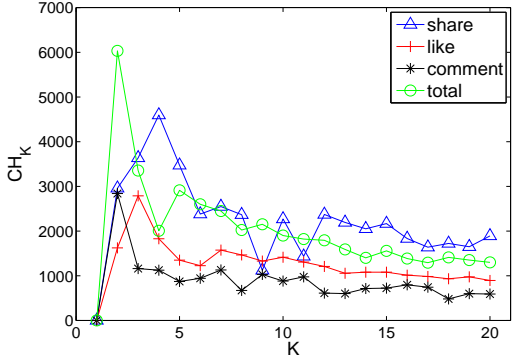

Figure 5: The variance of *CH-index* over cluster number $K$ of clustering *abnormal segment*. Higher CH-index indicates better clustering result.

*Feature vector.*

By far, each website can be represented by a set of abnormal segments. Intuitively, a website is apt to have a response pattern to a category of events rather than a particular one. Correspondingly, these patterns are better expressed by categories of segments than individual ones. Therefore we cluster all the segments across different websites into different categories.

The *abnormal segments* acquired are aligned to the same length for clustering. According to Assumption 1, the aligned length is set to $L$, which is 5. The aligning process is as follows.

- For an *abnormal segment* shorter than 5, the points preceding and following it in the *ratio track* are concatenated symmetrically to it, extending the length to 5. If this extension cannot be accomplished symmetrically, the preceding points are concatenated first.

- For an *abnormal segment* longer than 5, they are iteratively split into two parts at the lowest point until every new part's length is shorter than or equal to 5. Then the new parts shorter than 5 are extended as aforementioned.

After the alignment, the segments are clustered using K-medoids algorithm. The best number of clusters for each `X-count` is decided by the CH-method detailed in the next subsection. Fig. 5 illustrates which number of clusters works best. As a result, the `share count`, `like count`, `comment count` and `total count` are clustered into 4, 3, 2 and 2 categories, respectively. These 11 categories together constitute the feature vector of the websites. Each entry of the vector stores how many segments belonging to that category the website has.

### 3.3.3 Clustering algorithm

Without prior knowledge about the fake Like buyers, we are confronted with an unsupervised clustering problem. To solve this kind of problem, K-medoids and K-means are the most extensively applied algorithms. In contrast to K-means, K-medoids picks the medoid instead of the mean-value point as the cluster centroid in each iteration, and updates the new centroid by calculating pair-wise dissimilarity rather than using mean-value point of the intermediate clusters. Therefore, K-medoids is less sensitive to noises and outliers. In this paper, we choose the K-medoids algorithm.

When applying unsupervised clustering algorithm like K-medoids, an common problem is how to determine the best number of clusters. There are many techniques to address this problem [7, 18, 19, 25, 26]. In this paper, we choose the **CH method**, which is proposed by Calinski and Harabasz [7]. It is regarded as one of the most efficient techniques for determining the number of clusters [12]. The process goes as follows.

1. For a specific number of clusters $K$, cluster the data with a certain algorithm (K-medoids in our case). Then define two matrices $SSW$ (within-cluster scatter matrix) and $SSB$ (between-cluster scatter matrix) as follows,

$$SSW_K = \sum_{i=1}^{K} \sum_{\mathbf{x} \in C_i} (\mathbf{x} - \mathbf{m}_i)(\mathbf{x} - \mathbf{m}_i)^\top$$

$$SSB_K = \sum_{i=1}^{K} N_i (\mathbf{m}_i - \mathbf{m})(\mathbf{m}_i - \mathbf{m})^\top$$

where $C_i$ is the $i$th cluster; $\mathbf{m}_i$ is the center of $C_i$; $N_i$ is the number of points in $C_i$; $\mathbf{m}$ is the mean value of all points in the data set.

2. When having K clusters, the *CH-index* is defined as,

$$CH_K = \frac{trace(SSB_K)/(K-1)}{trace(SSW_K)/(N-K)}$$

where $N = \sum_{i=1}^{K} N_i$ is the total number of data points. Here the numerator describes the dispersion between the clusters, and the larger it is, the further different clusters are from each other. This means that the difference between these clusters are more obvious. In contrast, the denominator describes the dispersion within each cluster, and the smaller it is, the closer each element is to others in the same cluster. This indicates higher similarity among elements within the same cluster. Put them together, **higher CH-index indicates a better clustering result**.

3. Finally, find the best number of clusters, which is $\hat{K}$, the solution of the following optimization problem,

$$\hat{K} = \underset{K}{\operatorname{argmax}}\, CH_K, \quad K \geq 2$$

### 3.3.4 Clustering result

After the previous steps, each website can be represented by a 11-dimension feature vector. We notice that there are a lot of websites having all-zero feature vectors. These websites have no changes in either of the four counts during the observation period. Thus they can be confirmed as benign websites that never bought fake Likes. After excluding them, 3,483 websites remain to be clustered. The composition of these websites are given in Tab. 2.

By applying the CH-method, the best number of clusters is found to be 4 (Fig. 6). The clustered result is shown in Tab. 3.

Table 2: Composition of websites being clustered

| Groups | Top | Middle | Tail |
|---|---|---|---|
| Size | 1546 | 637 | 1300 |
| Percentage(%) | 44.8 | 18.3 | 37.3 |

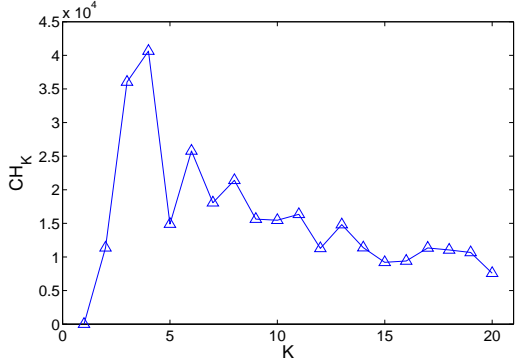

Figure 6: The variance of *CH-index* over cluster number $K$ of clustering the websites. Higher CH-index indicates better clustering result.

Table 3: Clusters of websites

| Clusters | 1 | 2 | 3 | 4 |
|---|---|---|---|---|
| Size | 655 | 1443 | 69 | 1316 |
| Percentage(%) | 18.8 | 41.4 | 2.0 | 37.8 |

## 3.4 Result analysis

According to Assumption 4, the top websites are less likely to buy fake Likes. Thus they can be used to identify the benign clusters. Denote the set of 3,483 clustered websites as $W$, we are interested in how they distribute over different clusters. We define the following three ratios to ease the analysis of the clustering result.

Given a set $J_n$, which comprises the first $n$ websites from either the top, the middle or the tail group, the inner-cluster ratio over cluster $C$ is,

$$I_{J_n,C} = \frac{|J_n \cap C|}{|J_n \cap W|}$$

Here, $|\cdot|$ denotes the cardinality of a set.
The inter-cluster ratio $O$ is,

$$O_C = \frac{|C|}{|W|}$$

The *tendency ratio* $R$ is defined as,

$$R_{J_n,C} = \frac{I_{J_n,C}}{O_C}$$

$R_{J_n,C}$ reflects the tendency of the websites from set $J_n$ appearing in cluster $C$. If the websites are uniformly distributed over the clusters, $R_{J_n,C}$ will always be 1 for any $C$ and $J_n$. The larger $R_{J_n,C}$ is, the more likely a website from set $J_n$ appears in cluster $C$.

### 3.4.1 Observations

Fig. 7 gives details about how the tendency ratio changes with different $n$, clusters and website groups. Each subfigure corresponds to one website group. From these figures, the following observations are made.

- Cluster 3 has the smallest tendency ratio for the top websites and largest tendency ratio for the tail websites. Particularly, none of the first 200 top websites appears in this cluster. Also, Cluster 3 has the smallest size (Tab. 3). According to Assumption 3 and 4, cluster 3 are most Likely to be the suspects of fake Like buyers.

- Cluster 4 has a dominant tendency ratio for the top websites. Besides, the leading sites (smaller $n$) in the top group correspond to a much higher tendency ratio than the rest ones. For example, for the first 200 top sites, a stable tendency ratio over 2 can be observed. Furthermore, it has the smallest tendency ratio over both the middle and tail sites. According to Assumption 4, cluster 4 is the most likely to be of benign websites.

- Cluster 1 and 2 come in between. Though the tendency ratios of cluster 1 for roughly the 50 leading sites in top group are zero, as $n$ increases, it quickly stabilizes at around 0.8, a higher level than cluster 2 and 3. Besides, its tendency ratios over both middle and tail sites are close to 1. Thus its elements can be treated as benign sites with high confidence. For cluster 2, its tendency ratio over the top and tail groups are only slightly better than cluster 3. However, we regard it as benign sites, for two reasons. First, it is the largest cluster comprising over 40% of the websites being clustered. If it is the suspect cluster, Assumption 3 will be false. Second, its tendency ratio over the middle and tail sites are consistent, this characteristic is more like cluster 1 rather than cluster 3, and implies that the tail sites are not special within the cluster.

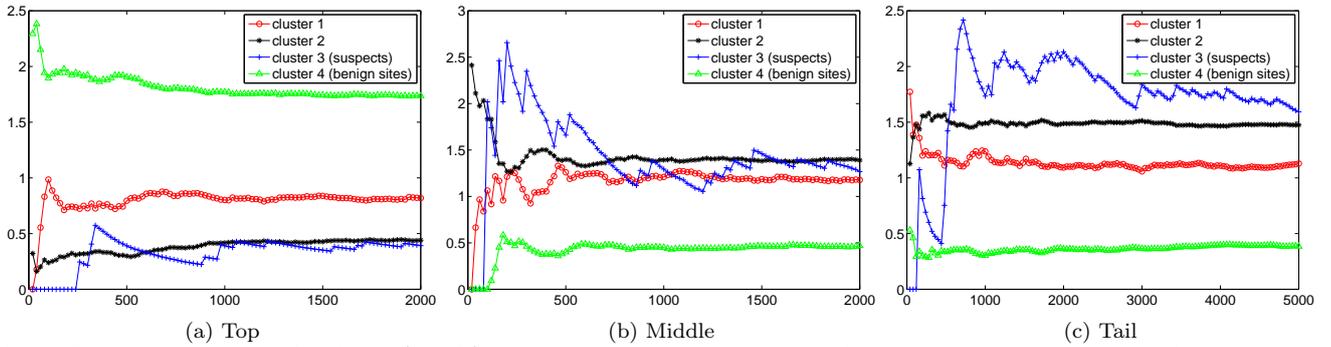

(a) Top    (b) Middle    (c) Tail

Figure 7: The tendency ratio $R_{J_n,C}$ (y-axis) of each cluster $C$ over the website set $J_n$. The set $J_n$ comprises the first $n$ (x-axis) websites from either the top, the middle or the tail group. If all the groups are uniformly distributed over different clusters, $R_{J_n,C}$ will always be 1. Higher $R_{J_n,C}$ indicates that the websites from set $J_n$ are more likely to appear in cluster $C$.

As a result, we conclude websites in cluster 3 to be suspects of fake Like buyers and those in cluster 4 to be confident benign websites.

### 3.4.2 Patterns of the websites

We manually inspect into the *ratio tracks* of both the suspicious and benign websites, and summarize the following patterns. Fig. 8 gives three websites' *ratio tracks* for illustration.

For the suspects of fake Like buyers:

- For most of the time, the curve of variance ratio stays stable at 0, which means no new Likes are generated at all.
- At some time point, the curve soars up and reaches a high ratio and then falls back dramatically, which forms a narrow spike in the graph. These spikes indicate potential fake Like generations.

On the contrary, for the benign websites:

- The *ratio track* keeps oscillating over time. These oscillations represent the daily changes of Likes at the normal level.
- At most of the time, the variance ratio is a small value above zero. Although there are peaks and valleys, they stay in a reasonable range.

### 3.5 Validation

First, we manually check the ratio tracks of the suspects and confirm that all of them express the characteristics aforementioned.

Second, to exclude the possibility that the high variance ratio is caused by small base number of `X-count`, we investigate the base number of counts for all the websites, which is listed in Tab. 4. From this table, we conclude that over 80% of the websites have the `X-count` smaller than 300. Thus, if a website has an absolute change larger than 300 in `X-count` for two consecutive records, it is reasonable to detect it as suspicious. We apply this "larger-than-300" constraint to the suspects in cluster 3 and have 16 of them remained. These 16 websites are highly suspicious and are regarded as our final detections of fake Like buyers. Specifically, the two suspects illustrated in Fig. 8 are among the 16.

Table 4: Base `X-count` of different websites

| Percent of websites (%) | | 50 | 80 | 90 | 100 |
|---|---|---|---|---|---|
| `share count` | ≤ | 5 | 128 | 1,211 | 3,064,140 |
| `like count` | ≤ | 1 | 82 | 665 | 19,991,431 |
| `comment count` | ≤ | 0 | 56 | 517 | 1,125,401 |
| `total count` | ≤ | 8 | 291 | 2,694 | 20,078,074 |

An alternative method is to look into the website's traffic to get an idea about whether the suspects had a dramatical uprise of traffic during the time period that they got an explosion in display count. However, this alternative method is not available directly, because the accurate traffic is confidential data of the websites that we as researchers can not access, and for those third party companies that provide estimated traffic data, such as Alexa, Compete, QuanteCast, etc., their estimation is not quite accurate. In some cases, the estimation can even have a bias off the real data for 2000% [9].

## 4. RELATED WORK

Detecting cheatings, fake accounts and spammers have been hot topics for years, due to the popularity and widely concern related to social networks. However, not much work related to the fake Likes exist. There are mainly two reasons: first, most researchers are not aware of the fake Like phenomenom and the black market behind it; second, it is difficult to obtain the ground truth about which Likes are fake and which websites are fake Like buyers. In this section, we elaborate upon the related research from two aspects.

### 4.1 Like button and the user privacy

As a platform for users to post daily updates, chat with real-world friends, share their geometrical locations, social networks involve much more private information than any other kind of traditional Internet services. Researchers have always been interested in how the social network companies handle these private information. A. Roosendaal first reveals that Facebook is tracking and tracing user information using the Like button and cookies behind it, and reaches

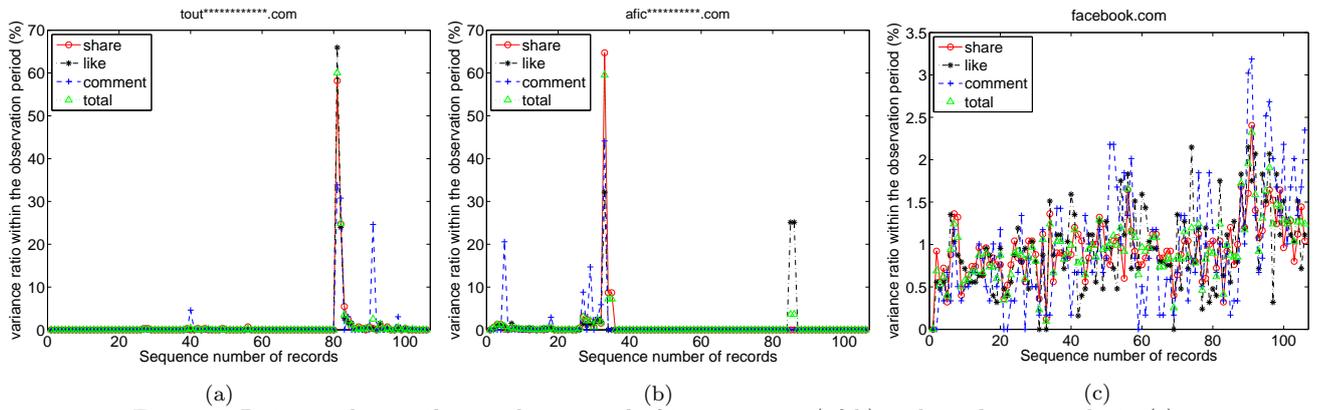
Figure 8: Ratio track over the crawling period of two suspects (a&b) and one benign website (c).

far beyond the Facebook platform to almost the whole Internet [21]. C. Wills and C. Tatar investigated how Google and Facebook employ the users' information to target the ads [31]. They observe that, if Facebook Like button is used by the users to express their interest, Facebook appears to target users with sensitive ads based on these interests. These works suggest that the Facebook Like related data are widely adopted for commercial use, which stimulate the needs of fake Likes.

### 4.2 Detect the fake accounts and spammers

With the rapid growth of social networks (e.g. Twitter, Facebook, Google+), the problem of detecting fake accounts is becoming a hot topic. Fake accounts and fake Likes both abuse the social networks for profits, and have intrinsic similarities in terms of spamming. Many techniques have been proposed to address this kind of problem.

**Analyzing the user profiles** is a major approach utilized by researchers. According to the official website of Facebook, there are at least ten ways to detect fake Facebook accounts by analyzing the user profile [23]. Similar approaches are applied to Twitter [3, 17, 28]. These approaches include checking the existence of the profile photo, investigating friend list (i.e., followers/followings), etc. These methods are easy to understand and implement, but when it comes to the fake Like problem, where these extra information is absent, they are not applicable anymore. H. Gao et al. focus on the spam of wall posts in Facebook [14], and manage to locate the spammers and characteristics. However, all these work are trying to fight against fake/spam accounts, which cannot directly apply to our problem, because the intrinsic differences between fake Like buyers and spammers. Rather than spreading any spams or taking hazardous actions towards other users, the fake Like buyers are hiding in the dark and stimulating the spammers and fake account owners to do so, and then benefit from these malicious behaviours. Thus they are not likely to be detected by utilizing the same methods of detecting fake/spam accounts.

**Machine learning algorithms** also play an important role in detecting fake accounts and spammers. For instance, SVM [3, 10, 15] are used for spammer detection on Twitter, and Naive Bayes [2] classification algorithms were used for email spam categorization. These choices are made because it is assumed that obtaining the ground truth is easy, with the help of a human being. However, this assumption no longer holds in the case of fake Like detection. Due to the lack of related information, even a human being cannot tell a real Like from a fake Like or a benign website from a fake Like buyer. That's also the reason that we choose unsupervised learning method in our work, taking use of K-medoids algorithm.

**Honeypot** is another strategy widely used to detect spammers [15, 24, 29] on social networks. The idea of honeypot is straightforward: the researchers first analyze the targets which are attractive to the spammers, and extract the common features of these targets; then they use these features to setup bait profiles, which will attract the spammers to attack. These profiles are the "honey"; finally, they release these "honey" profiles to public, then lure and capture the spammers. The challenging part of applying the honeypot strategy to the fake Like problem is, the actual buyers are hidden behind the fake Like spammers. Even if we successfully catch the spammers, it will be difficult to distinguish whether they are randomly spamming or paid to do so.

## 5. CONCLUSION AND FUTURE WORK

In this paper, we conducted extensive research on the Facebook Likes. We demonstrated that the design and implementation of the Facebook Like not only violated the natural meaning of "like" but also left behind deficiencies that can be utilized to generate fake Likes. We discussed possible ways to generate fake Likes and as an example, successfully generated 100 fake Likes within 5 minutes, using a single account. By applying K-medoids clustering, we found out suspects of fake buyers from the real data set crawled from Facebook, and concluded the basic characteristics of these suspects.

Our work leads to many interesting yet promising research topics. After figuring out possible fake Like generating methods and finding out potential fake Like buyers, how to locate and eliminate these fake Likes remains an open question. As we suggested, a desgin with full accountability is a potential solution. Besides, it is also necessary to dig out how and how much a website may benefit from fake Likes to help achieve a better understanding of the fake Like black market. Furthermore, other social networks also have similar features, such as "+1" of Google+, and "retweet" for Twitter. A research on these features may be compared to our current work to get a comprehensive understanding of this kind of problems in social networks. We leave these topics as future work.